\documentclass[a4paper,12pt]{article}
\usepackage[left=2.5cm,right=2.5cm,
top=3cm,bottom=3cm]{geometry}
\usepackage[T2A]{fontenc}
\usepackage[utf8]{inputenc}
\usepackage[english]{babel}
\usepackage{amsmath,amssymb,amsthm,mathtools,bm}
\usepackage{euscript,mathrsfs}
\usepackage{icomma}
\usepackage{graphicx}
\usepackage{wrapfig}
\usepackage[dvipsnames]{xcolor}
\usepackage{indentfirst}
\usepackage{amsbsy}
\usepackage{wasysym}
\usepackage{hyperref}
\usepackage{cancel}
\usepackage{verbatim}
\usepackage{empheq}
\usepackage{adjustbox}
\usepackage{authblk}
\usepackage{tikz-cd}

\begin{document}

\title{ Consciousness via MIPT?  }
\author{Alexander Gorsky$^{1,2}$ }

\affil{$^1$Institute for Information Transmission Problems RAS, 127051 Moscow, Russia \\ 
$^2$Laboratory of Complex Networks, Center for Neurophysics and Neuromorphic Technologies, Moscow, Russia
}

\maketitle
\begin{center}
 {\bf Abstract}   
\end{center}
The measurement-induced phase transition (MIPT) is a recently formulated phenomenon in out-of-equilibrium systems. The competition between unitary evolutions and measurement-induced non-unitaries leads to the transition between the entangled and disentangled phases at some critical measurement rate. We conjecture that self-organized MIPT plays a key role in the generative model of cognitive networks and the formation of the state of consciousness in the "newborn-adult" transition. To this aim, we formulate the probe-target picture for the brain and suggest that MIPT interpreted as learnability transition takes place in the mental part of the target where the
sites in the cognitive networks of semantic memory are concepts. Comparison with the synchronization phase transitions in the probe is made.



\maketitle
\vspace{3cm}
\section{Introduction}

The physics of out-of-equilibrium phenomena provides a possible framework for brain studies; see \cite{nartallo2025nonequilibrium} for a review. 
Among them, the possible near-criticality of the operating state of the brain is considered as attractive hypothesis
\cite{mora2011biological,bak, pikovsky2001synchronization,buzsaki2006rhythms,munoz2018colloquium,plenz2021self}.
A more complicated question is
whether some version of criticality is relevant for the generative model of consciousness. 
In \cite{tegmark2015consciousness}, it was suggested that consciousness is the special state of matter that can emerge through a phase transition, and some aspects of the relevance of phase transitions to consciousness were discussed in \cite{kozma2018phase,chialvo2010emergent,crick2003framework}. The attempt to apply quantum mechanics and formulate a hypothetical relation between consciousness and the collapse of the wave function was made in \cite{hameroff2014consciousness}. The two most popular "theories of consciousness" - Integrated Information Theory \cite{tononi2004information,tononi2015consciousness} and
Global Neuronal Space Theory \cite{dehaene2001towards} - do not focus on the generative models.
The review of current theories of consciousness and their classifications is presented in \cite{seth2022theories}, while a discussion of the transition time schedule from newborn to adult can be found in \cite{bayne2023consciousness}.

In this paper, we shall focus on the possible application of MIPT in a monitored out-of-equilibrium system formulated in \cite{li2018quantum,skinner2019measurement,chan2019unitary}, see \cite{fisher2023random} for a review, to the generative model of consciousness. The idea behind the MIPT is quite simple; we have an open system interacting with the probe device providing measurements. The interplay between the unitary evolution and the nonunitarity induced by the monitoring amounts to the phase transition or crossover at some critical measurement rate. The transition occurs between the
entangled and disentangled phases, and the behavior of the entanglement entropy serves as a possible order parameter. 
This was the initial interpretation of MIPT.
Later, a slightly different interpretation of MIPT has been developed as a learnability transition \cite{Barratt_2022,Ippoliti,agrawal2024observing}.  In the entanglement phase at a weak measurement rate, the initial information is hardly extracted, whereas at a strong measurement regime the monitoring device can extract information about the initial state of the system. In this interpretation,
the order parameter for the transition is the Shannon entropy of the measurement record \cite{PhysRevLett.125.030505}.

It was argued that in the generic situation the monitoring process involves two parameters. The first parameter P quantifies the frequency or probability of the measurement, while the second parameter M quantifies the strength of the measurement. The phase structure for MIPT in the (M,P) plane has been developed in
\cite{szyniszewski2019entanglement, szyniszewski2020universality}. In addition, the interesting effect of measurement has been found in \cite{doggen2022generalized}, which will be important in our study. It was shown
that there is a domain on the (M,P) plane when the measurement induces the attraction between the constituents and their clusterization. This
domain corresponds to the monitoring regime when the parameters M and P are approximately of the same importance.

We conjecture that MIPT is an important mechanism that contributes to the formation of the consciousness state. The application of MIPT to the generative model for the consciousness requires at least qualitative answers to a few immediate questions. 
\begin{itemize}
    \item What is the target system under measurement and what is the probe monitoring device?
    \item What is a single act of measurement?
    \item What is the order parameter for MIPT in the brain?
\end{itemize}

To approach this issue, we formulate the "probe-target" perspective partially inspired by the holographic principle \cite{hooft1993dimensional,susskind1995world}.
Both probe and target are structures that develop under a generative process. The probe evolution starts with the structural connectome, and eventually the ensemble of the interacting functional networks is formed. The target evolution
starts with the ensemble of raw subjective experiences-qualia that have some physical carriers in the synaptic connections and are simultaneously elements of the abstract mental space \footnote{Another used notion for this object - subjective space}. Measurements induce the interaction between qualia in the mental space in the proper domain in the (M,P) parameter plane near MIPT. Due to effective attraction, they form the cluster of qualia of a similar context, called a concept following \cite{lewis1956mind} and which was extensively discussed in \cite{rusakov-concepts}. Eventually, a set of concepts forms
the mental component of the target, that is, the semantic memory organized
as the hierarchical network. We suggest that the recollection of the selected qualia is the unit measurement act.

In this picture, we view the physical brain as a measuring (learning) device that creates the mental part of the target
manifold and continuously monitors elements of the target to gain information.
The order parameter for this learnability transition is the Shannon entropy of the measurement output. To some extent, our picture refines the "consciousness as the collapse of the wave function" scenario in \cite{hameroff2014consciousness} by a much softer
picture of the formation of the network of semantic memory by the MIPT which "resolves" the wave function collapse. 
Although we used intensively the notion of entanglement entropy, we did not refer to quantum mechanics at all. The network
structure both in the probe and in the target provides the eigenvalue problem for the large N matrix, which
substitutes the eigenvalue problem for the quantum Hamiltonian. In some non-rigorous sense, $1/N$ plays the role of the
Planck constant in such problems described by large $N$ matrix models. Only probabilistic aspects of monitoring are important.

Let us comment on the previous literature. A kind of probe-target picture and the network approach to both of them have been partially discussed before. In cite \cite{Tegmark_2000} it was suggested
to work with the subject-object-environment picture to deal with the effects of decrease and growth of entropy. However, the definitions of object and subject
were not related to functional networks and concepts. Moreover, at that time the very idea of MIPT had not been developed. In
cognitome picture \cite{anokhin2021cognitome}  the whole brain-consciousness complex has been considered as the complicated network
where the single functional system unified with the corresponding stimulus forms the "unit of consciousness"- named "cog", the cogs are connected by links and
form the general hierarchical network. There is no separation between probe and target in the cognitome picture, and it is difficult to define the phase
transitions separately in the components, which we shall focus on.

Identification of qualia and measurement have been suggested in \cite{resende2022qualia} where possible mathematical aspects of the space of qualia have been elaborated. 
This mathematical structure implies that the concepts are formed in response to qualia while the concepts enable the existence of qualia.
It was assumed that the qualia belong to the mental space while the concepts are stored in the synaptic connections that they have physical
carriers. The separation of "internal" and "mental" degrees of freedom in \cite{resende2022qualia} is different from the picture we shall advocate in this paper.

The paper is organized as follows. In Section 2 we formulate the probe-target picture for the brain and the meaning of the individual measurement act. In Section 3 we recall the key elements regarding MIPT and formulate our conjecture concerning the generation of concepts from the qualia via the attraction provided by monitoring and formation of the
semantic memory via MIPT treated as the learnability transition. In Section 4 we compare the MIPT in the target and the relevant
synchronization phase transitions for the probe. In Conclusion we summarize our findings and formulate the open questions. A few
comments concerning the used notions can be found in the Appendix.

\section{Probe-Target picture for the brain}
\subsection{Definitions of probe and target for brain}

What are the probe and target space in the brain study? Let us start with the probe which
from the general viewpoint corresponds to the physical brain. However, we have to be more precise and consider two components of the probe. The first component is an underlying "neural manifold" that involves neurons physically organized into the structural connectome \cite{sporns2016networks,lynn2019physics}, which is the particular single network. It is now well established  that neurons are highly specialized, so they fire on particular external stimuli. A set of neurons located at different anatomical areas of the brain, while having the same specialization, will be called a "functional network" throughout this paper. \footnote{ Another term introduced in the same context a long time ago is functional systems- "Complexes of co-active elements of distributed anatomical localization that co-operate towards common adaptive result by the whole organism" (P.K.Anokhin, 1937)}. 
There is a functional network for each stimulus; some of them are "charged" with respect to the space, like place cells or grid cells \cite{moser2008place}, while some of them are "charged"
with respect to abstract stimuli like in \cite{bowers2009biological}. There is a recent study \cite{rey2025lack}
that confirms the context independence of the specialization of individual neurons charged with respect to the abstract stimulus. This feature seems
to be absent for other species.
Generally, there are an infinite number of functional networks. Moreover, some neurons are multi-specialized and thus participate in several functional networks simultaneously. This ensemble of interconnected functional networks is one component of our "probe ". The important fact is that neurons can change their specializations. 

The second component of the probe is episodic memory which has a physical carrier and is attributed at least partially to details of synaptic plasticity. The interplay between components and the timing of the flow from short-term to long-term memory is quite complicated.  In summary, we assume

$$\text {Probe = ensemble of interacting functional networks of neurons + episodic memory}$$

To describe the target manifold, where the probe "propagates in" or is mapped onto, we use the standard term "stimuli", which means any external signal to which a particular functional network reacts. There are two components of the target manifold. The first component is the material environment of the brain. The second mental component is the cognitive network, or equivalently, semantic memory. The nodes of the cognitive network
are concepts that form in some version of generative models from the unsorted ensemble of subjective experience -"qualia".
Note that the terms "qualia" and "concept" were first introduced in \cite{lewis1956mind} in the context of the theory of knowledge. The concepts were discussed in \cite{rusakov-concepts,rusakov-24} as a complete set of fundamental constituents of consciousness which presumably emerge through a phase transition. 
Semantic memory \cite{Saumier,Tulving_1984} can be considered as a hierarchical network in which at the lowest levels there are unsorted qualia, while at higher levels there are concepts eventually formed. 

Therefore, we define the target manifold as  
 
 $$ \text{target= ensemble of stimuli of any nature}$$
 $$ \text{ mental part of target = semantic memory}$$

 \begin{figure}
    \centering
    \includegraphics[width= \linewidth]{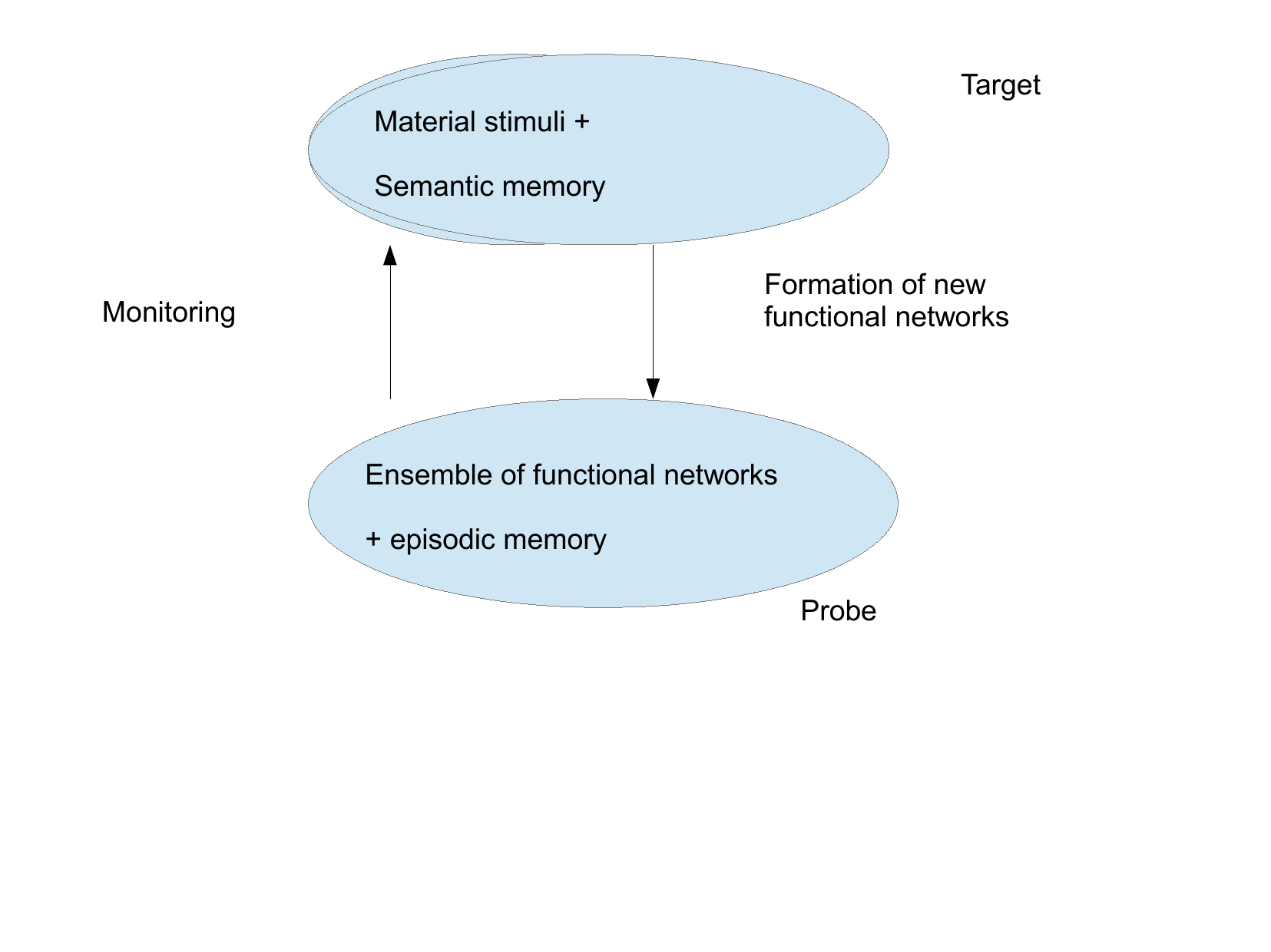}
    \caption{Schematic representation of the target-probe picture. The probe monitors the target while the
    emergence of new concept in the semantic memory induces the appearance of new functional network in the probe}
    \label{fig:MIPT3}
\end{figure}

There is a question concerning one aspect of the stimuli space that causes a controversy: Is space part of the target, just like abstract mental stimuli? In favor of this idea, it is spoken of the fact that there are place cells in the hippocampus and grid cells in the entorhinal cortex that are "charged" with respect to space and provide effective space navigation. These facts might suggest including space in the target on the same footing as the mental component. However, there is no consensus on this point in the literature (see \cite{buzsaki2017space,hameroff2014consciousness} for a recent review paper). In particular, recent studies of grid cells suggest that their
role can be more involved \cite{ginosar2023grid}.
The term semantic memory will be used for the mental part of the stimuli space.

We emphasize that while the probe is roughly equivalent to the physical brain, the target is a higher-dimensional (potentially infinite-dimensional) space associated with it. Both probe and target at least partially have the multilayer network structure, but with very different nodes and links.
Both exhibit a kind of plasticity, the neuron specialization can change, and the new functional networks on the probe can be formed. On the other hand, semantic memory can be enlarged due to new qualia and concepts.
We emphasize that our definition of the physical probe and the mental target is different from the related discussion in \cite{resende2022qualia}. In that paper, it was assumed that the concept is attributed to the physical brain and corresponds to the details of the synaptic connections.
In our picture, the concept is the node in the cognitive network in the target and its fingerprint on the probe is the corresponding functional network.

In the generic holographic framework the self-consistency of the probe-bulk picture imply the highly nontrivial constraints on their
interaction because of the possible back-reaction of the probe on the target. If we consider the interplay of the probe with the material stimuli, such a back-reaction cannot be expected. However, the interaction with the mental part of the target certainly involves the back-reaction; however, we are not able to completely describe it and shall focus in what follows on some simplified version.
There are at least two different situations.
The first pattern of interaction suggests that the new situation involving a new stimulus or new combinations of stimuli is on the scene, and the probe produces new qualia by coherent reaction of a particular combination of functional networks.
The second type of interaction, which we call recollection, is common in the process of thinking.
It is a kind of backward-forward process when the episodic memory part of the probe is involved. It is necessary to address episodic memory by taking off some item, "measure" it, and bring it to the target upon measurement. No new qualia are formed in this process, but this measurement act certainly influences the target providing the nonunitary evolution of the ensemble of qualia. In what follows, this process will be important for MIPT.

\subsection{What types of phase transitions are relevant?}

Let us make general comments on the proper types of phase transitions relevant for the probe and the target which we shall discuss in the next Sections. In the probe,
the key mechanism is the entanglement and consolidation of degrees of freedom.

\begin{itemize}
    \item {\bf Entanglement of components in the probe network via synchronization}.  At the micro level, the strongly correlated functional networks get formed from the weakly correlated ones via synchronization , thus leading to the formation of the global neuronal space \cite{dehaene2001towards}. In this  synchronization transition, the firing of individual neurons is not as relevant as the collective behavior of functional networks, which changes and produces the change in the band of the brain rhythms. This implies a sort of global synchronization of the networks. The interaction
    of functional networks can be via direct links or 
    by the effective  "entropic forces" which also do the job and produce the consolidation of components via Page transition, which has been discussed in this context in \cite{gorsky2022page}.   
    
    Additional partial synchronization of probe functional networks occurs as a response to stimuli. This transition is produced by the correlation of excitations of several functional networks on top of global synchronization. In the generic theory of synchronization, it would mean a formation of the chimera-like states.
    The simplest paradigmatic model is the Kuramoto model when the synchronization occurs due to the real interaction between the constituents \cite{pikovsky2001synchronization} and the phase transition with the formation of chimeras occurs at the critical coupling constant. 

    Remark that the architecture of the structural connectome matters and influences the pattern of synchronization. The key feature of the structural connectome relevant for synchronization is the Anderson localization of eigenmodes of the Laplacian network. It tells if the excitations are freely propagate the connectome or localized at some domains. Quite surprisingly, for the human connectome with large-scale resolution, it was found that the architecture implies the intermediate regime for the propagation of excitation between the completely chaotic and completely localized regimes \cite{pospelov2019spectral,bobyleva2025metric}.

    \item {\bf Disentanglement transition in the target } 

    In the mental part of the target the key question concerns the formation of the cognitive network or semantic memory from the entangled unsorted qualia. The underlying mechanism is opposite to the probe - the disentanglement of the initially entangled degrees of freedom. 
    The formation of the semantic memory as the hierarchical cognitive network
    is the complicated generative process.
    The elementary transition involves the formation of the single concept from ensemble of qualia  however
    at this stage no general formation of the cognitive network takes place.
    We suggest that the underlying mechanism for formation of concept from qualia is the effective attraction 
    of monitored constituents. 
    The effective disentanglement of group of qualia from the rest takes place and a new site of the future cognitive network gets formed. 
    Increasing the measurement rate more concepts get formed and eventually the cognitive network gets emerged,
The important feature of the MIPT is the change in the entanglement entropy pattern from the "area law" to the "volume law" in the object of measurement. Presumably, this transition can also be related to the induced information geometry in the target space. It is a rather common situation where extra dimension emerges in the target manifold at the point of singularity. We shall discuss
the conjectural role of MIPT in formation of semantic memory in the next section.

\end{itemize}

\section{Formation of cognitive network via MIPT. Newborn - adult transition}
In this Section we shall discuss the generative model for the semantic networks in the mental part of the target manifold. which is the central aim of our study. We suggest
that MIPT is an underlying mechanism for the formation of sites in the mental part of the target and the target network itself.

\subsection{Generalities about MIPT. Single control parameter}

The MIPT involves two ingredients - unitary evolution and the nonunitaries due to the monitoring the system via measurements. The systems under consideration can be
different, like spin chains, the monitoring can be periodic or not, there can be some number of additional conservation laws, etc. Initially, MIPT was related to the change in the
entanglement pattern from the volume law in the weakly monitored system to the area law for the entanglement entropy at high measurement rate \cite{li2018quantum,skinner2019measurement,chan2019unitary}. However, later other
important interpretations have been suggested.  MIPT was related to the purification transition when the purification rate of the initially mixed state was changed \cite{gullans2020dynamical}. The behavior in the volume-law regime has been related to error-correcting codes \cite{choi2020quantum,fan2021self}. Finally, the MIPT transition has been related to the learnability
transition \cite{Barratt_2022,agrawal2024observing,Ippoliti}, that is, the measurement device can learn the
initial state of the system at a high measurement rate while it cannot effectively learn at a low rate. The optimal learning corresponds to the measurement rate in the transition domain.

First, recall the interpretation of MIPT as a transition
in entanglement.
The entanglement entropy measures how the statistical entropy of measurement of subsystem A is reduced if one first completely measures
the state of B. For pure state reduced density matrix reads as 
\begin{equation}
    \rho=|\Psi><\Psi| ,  \qquad \rho_A=Tr_B\rho  
\end{equation}
The corresponding Renyi entropies are defined as 
\begin{equation}
    S_n= \frac{\log Tr(\rho_A)^n}{1-n}
\end{equation}
which for $n=1$ corresponds to the von Neumann entanglement entropy. If each site in the system involves q degrees of freedom $S_n\rightarrow S_0$ for all
values of n.

In the entanglement phase $p<p_c$ $$S(t, L\rightarrow \infty  ) \propto t, \qquad   S(t\rightarrow \infty,L  ) \propto Volume  $$ while
in the disentangled phase at $p>p_c$ 
$$S(t, L\rightarrow \infty  ) \propto const, \qquad   S(t\rightarrow \infty,L  ) \propto Area  $$ 
in (1+1) the area of the boundary is just constant. In more informal terms in the entanglement phase, the entanglement grows linearly in time and saturates at a value
proportional to the subsystem size. In the disentangled phase, the initial state with low entanglement never becomes extensive. The growth rate tends to zero at O(1)
time scale. If the initial state is highly entangled at critical $p=p_c$ its entanglement drops to the O(1) value. At $p=p_c$ we have critical dynamics with
the corresponding critical indices.

The microscopic picture behind the 
entanglement approach to MIPT uses several ideas. First, we mention the membrane paradigm, which is quite intuitive \cite{zhou2020entanglement}. If we divide the system
into two parts, A and B, the curve connecting the boundary between A and B at initial and final times is a natural object. At $q\rightarrow \infty $ it corresponds
to the minimal cut minimizing the entropy, while at finite q both the "energy" and "entropy" contributions must be taken into account when considering the membrane surface.
The membrane tension is subject to renormalization which affects the point of the phase transition. In the most intuitive way, the
renormalization is seen in the effective spin system such as Ising for the measurement \cite{jian2020measurement,bao2020theory}. Then it goes as follows. Consider the forward and backward
unitaries $U$ and $U^*$ which for the density matrix form a structure $U^*\rho U$ which after some massaging can be reformulated
as the Ising-like spin degree of freedom $\sigma$ taking values $\pm$ and the orthogonal component $\sigma_{\perp}$ . 

In the entanglement phase, the membrane can be considered as the Ising model domain wall that separates the $+$ and $-$ phases. However, the domain wall prefers
to go through the bubbles of the $\sigma_{\perp}$ and crossing these domains renormalizes the
membrane tension since the cost function in this domain is different. When the Ising chain gets monitored, the measurements that drop some
links in the unitary evolution induce the disorder in the 2d effective spin chain model and at some critical disorder the drops in the unitary evolution start to dominate.

The membrane picture allowed us to relate the problem to the dynamics of the directed polymer in the background of the
obstacles. The polymer picture provides the additional tools to analyze the problem and in particular
the KPZ scaling for the entanglement entropy in the entanglement phase has been predicted and confirmed numerically in (1+1)
\begin{equation}
    S(x,t)=v_e t +t^{1/3}\chi
\end{equation}
where $\chi$ is the random variable.

The interpretation of MIPT as the purification transition
goes as follows \cite{gullans2020dynamical}. The measure of purification
is the transition rate from mixed state to pure state.
If we start with the strongly entangled mixed state and start to monitor the system at the measurement rate $p>p_C$ the fast purification phase occurs at some time. The mechanism of purification is as follows. At some time $t^*$ the number of "holes" induced by measurements in the evolution networks becomes large enough and the "percolation through the holes" at some time becomes possible separating the state of the system at $t<t^*$ and $t>t^*$ by an effective membrane extended in some time interval \cite{gullans2020dynamical}. At $t>t^*$ the system becomes almost pure quickly with well-separated constituents that can be approximated by the state of the matrix product. To some extent, this transition can be interpreted as an example of dynamic phase transition \cite{heyl2018dynamical,zvyagin2016dynamical}.
\begin{figure}
    \centering
    \includegraphics [width=0.8 \linewidth]{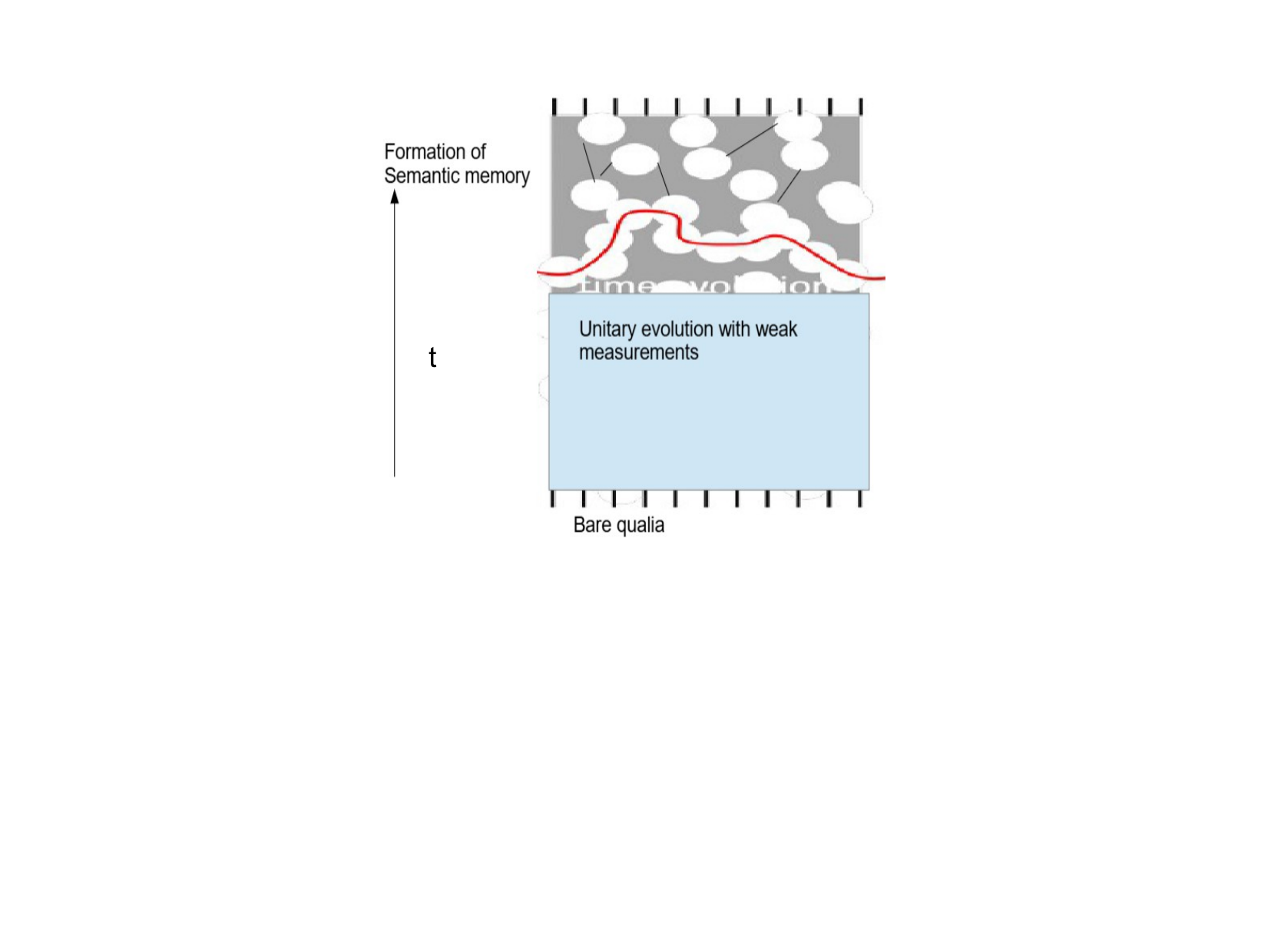}
    \caption{Schematic representation of the MIPT as dynamical transition in some time interval }
    \label{fig:MIPT1}
\end{figure}
The third learnability transition interpretation of MIPT is most interesting for our purpose. It has been argued \cite{Barratt_2022,agrawal2024observing,Ippoliti} that in the high measurement rate regime the measurement device learns and can obtain the initial state of the evolving monitored system. It becomes even more transparent when there is at least one conserved charge in this system. In the low-measurement phase, the chaotic background hides the initial information from measurement. Importantly, the learnability approach uncovers that the transition can be detected without looking at the particular post-selected states. In this perspective, the
order parameter for this transition is different and can
be extracted from the correlations of the measurement outcomes.

\subsection{Two control parameters}

In a more general situation which is relevant for our study, two control parameters are introduced.
To describe the generic monitoring of the system, let us consider the two-dimensional parameter space $(M,P)$ where
M measures the frequency of the measurements or the probability of the measurement while P measures the strength of the measurement.
In this case, the phase structure is more rich and new interesting phenomena can be recognized. 
It turned out that the several universality classes are involved in the phase diagram. There are the percolation
critical point at large $P_c$ and the point at Nishimori line connected by the crossover curve in the $(M,P)$ 
plane \cite{pütz2025flownishimoriuniversalityweakly, zabalo2022operator}. In the weakly monitored case the Nishimori line
corresponds to the stability domain in the renormalization flow \cite{pütz2025flownishimoriuniversalityweakly} and the crossover line separates two phases with the different entanglement patterns. Two control parameters correspond to two different ways of introducing non-unitarity in the system.

Another interesting phenomenon for the model with two control parameters has been found in \cite{doggen2022generalized}.
It was shown that there is the domain in the parameter plane where
the attractive interaction between the constituents induced by measurements is generated near the phase transition.
In \cite{doggen2022generalized} it was argued that attraction-induced clusterization takes place in the domain of the parameter plane $(M,P)$ when M is large enough
while P is in the intermediate regime. Moreover, it was shown that the clusterization takes place when the effect of unitary
evolution is compared with the non-unitary one. The induced attractive interaction takes place even if there is no
bare interaction between the constituents.

The measurements induce the disorder into the system, hence one could question the fractality in the properly defined
moments of the operators. It turns out a bit surprisingly that near the crossover line the multifractality of the spectrum
of entropic variables was identified \cite{pütz2025flownishimoriuniversalityweakly, zabalo2022operator}. Multifractality
survives even in the percolation limit.

\subsection{MIPT in the brain}

Have we any ingredients of the MIPT in our case? We first have to recognize the very system that we are looking at, the measurement unit, etc.
Assume that you are a newborn - no concepts in the target and therefore no corresponding functional networks in the probe. Nevertheless, you start to monitor the world around by available means, and the qualia or episodic memory unit gets generated. At some moment of time, you have a large enough ensemble of qualia, and we start investigating its time evolution. Let us emphasize the specifics of our case - we generate the system under consideration via measurements since the set of qualia is our system under study. In the conventional MIPT, the system is prepared before. At the initial period of time we are in the unstructured entangled phase of the ensemble of qualia, and the system tends to "thermalize".

Time is passing and you are a bit older. You have a large ensemble of qualia, and two operations are possible. First, you can get a new subjective experience and therefore add new qualia enlarging this dimension of the space under consideration. Secondly, you can intensively monitor the set of existing qualia with some measurement rate $p$. Each such measurement
involves the backward move "recalling of this qualia" resembling the forward-backward coupled pair which is central in the Ising-like description of MIPT.

We can expect at least two properties for the system of qualia. First, we should expect some mechanism to provide
their effective "attraction". Secondly, we expect that the measurement could recover the initial state, that is,
the individual qualia. In fact, we can recall our
individual experiences. Hence, we have to be in the phase of high learnability. 

Hopefully both ingredients are provided by MIPT. 
First, note that each measurement induces a kind of conservation law which is the unit of memory. 
Hence from the formal viewpoint we consider the system with emerging conservation laws, that is, emerging symmetries responsible for these laws. These emerging symmetries can be approximate and unstable if the qualia does not propagate into long-term memory or almost exact when it does. 
The presence of the effective conservation laws justifies
the learnability transition viewpoint since just in this situation at high measurement rate device-probe learns the initial states of constituents.

Now we have to explain how the individual concept is formed from qualia and how the network of concepts is eventually formed. We assume that unsorted qualia correspond to the entangled phase of the mental part or the target and evolve approximately unitarily. The measurements
yield non-unitarities in the qualia evolution; however, for the initial period of time the measurement rate is subcritical. If the baby starts to
attempt to classify the world's items of a related nature, he/she addresses the corresponding related qualia. The measurement act is
identified with the recalling of qualia - it produces nonunitarity. The key point is that nonunitarity can induce the interaction between
evolving elements \cite{doggen2022generalized} and eventually results in clusterization. At some baby age the measurements become intensive enough
and the individual clusters in the mental part of target manifold involving the qualia of a similar context emerge because of the
interaction induced by nonunitarities. The individual concept gets formed. It seems that the formation of the single concept is a
short-time phenomenon and the whole mental target is still in the entangled phase.

However, the single concept cannot rearrange the whole phase
of the mental manifold, it is local. To get the new global phase the MIPT should occur. It could happen at some critical
rate of recollecting related qualia of different contexts. Multiple concepts are formed, and MIPT in the
network framework corresponds to the clusterization phase transition with a macroscopic number of concepts. Generically, there are different ways to induce the clusterization transition in the abstract networks,
but here it is induced by monitoring the target space by the probe. The timescale for the rearranging of the whole mental target
is different, and the time interval is much longer than the time to form the individual concept. Hence, to some extent we have the Griffiths-type phenomenon with an extended phase transition or the crossover. There is a similarity with the Griffiths phenomenon
indeed, since here we have formation of the bubbles of the new phase, eventually forming the new phase via interaction. This is
the standard framework in this context.

This scenario provides possible answers to the questions disputed in the context of the newborn-adult transition \cite{bayne2023consciousness}.
Is the transition a short-term process and do all attributes of consciousness appear more or less simultaneously? In our approach, it is clear
that the process requires the critical rate of measurement during some time period, and formation of different concepts and links between concepts does not occur simultaneously.

Note the special role of episodic memory in this approach. The qualia
simultaneously are present in the target where they are the bare objects which the concepts are eventually built from, and
in the probe presumably located in the synaptic connections. In the target, they get clustered into units of semantic memory via MIPT.
In the probe the mechanism which yields the new functional network specialized in a new concept
from the ensemble of episodic memory elements is unclear.

We emphasize that MIPT is induced by the monitoring of the mental part of the target,
which also occurs at different rates at sleep. The monitoring of the material part of the target
can influence the MIPT at awareness state due to interaction of two types of functional network on the probe.
The monitoring could involve additional ingredients like respecting some symmetries, and, in general, we could assume that the monitoring could involve a kind of optimization
of the architecture of cognitive networks. Some guiding rules, such as the free energy principle
\cite{friston2010free,dacosta2024mathematicalperspectiveneurophenomenology,whyte2024minimaltheoryconsciousnessimplicit} can be relevant for the measurement item.

If we do not assume the existence of a "center of monitoring" at sleep, it could be natural to suggest that
the MIPT could be the version of self-organized criticality in the
context of generative model for consciousness. That is, the measurement strength and intensity are adjusted in such a way that
the target cognitive networks are close to the MIPT regime.
Some attempts have been made to develop the
self-organized criticality approach for MIPT in the peculiar model in condensed matter \cite{fan2021self}.
It was also argued that in the learnability transition,
which we identify with the MIPT the very transition
critical parameter corresponds to the optimal learnability.
Although this idea is attractive, we cannot suggest its precise realization at the moment.

Can the membrane -like picture of MIPT be applied to the formation of semantic memory and therefore conscious state?
Assume that the measurement rate is large enough $p>p_c$ and we fall into the disentangled regime. 
We can speculate that the membrane-like structure gets formed from the measurements involving qualia with related context, which eventually form a trajectory
in the qualia space with some tension. This trajectory can form a closed curve with the disentangled droplet inside.
We try to identify such a bubble with the formation of a single concept, and the boundary of the bubble is formed from the chain of qualia related to this concept. The tension can be equalized by the difference in the entropic pressure in two phases. However, such a qualitative droplet picture cannot be considered without additional supporting arguments.

There are no doubts that it is an oversimplified picture, but it could provide a proper starting point.
Such periodic monitoring of the target by the probe induces an additional rearrangement of the entanglement in the
target manifold. For example, new links between concepts can be created.  This multiple
measurement sequence eventually results in the $newborn \rightarrow adult$ transition, which we consider as the basic example
for the transition to the conscious state. Our arguments above
suggest that it is probably a transition of Griffiths type.

There is an evident naive analogy between the $newborn\rightarrow adult$ transition and the $animal\rightarrow human$ transition,
and the question whether the MIPT is relevant in any sense for the second transition is meaningful. To answer this question, we have to attempt
to define the analogue of the single qualia and the single measurement. We could speculate that the analogue of the qualia
is the mutation and the analogue of the measurement is the estimate of the positivity of the evolutionary impact
of the particular mutation in the spirit of the principle of free energy applied in the evolutionary context \cite{friston2010free}.
The mutations evolve unitarily while the impact estimation induces the formal nonunitaries. Hence, indeed, we are in the framework
of MIPT. As before, there are two parameters (frequency,intensity) that quantify the effects of measurement.
It is clear that the measurements and mutations are much slower than in the $newborn \rightarrow adult$ case, and therefore the dynamical
crossover to the human is very slow.

\subsection{Comment on the transitions in cognitive networks}

We have discussed multiple functional networks on top of the same structural connectome
on the probe. Can we similarly consider the " structural architecture" and the "functional
cognitive networks" in the target?
With the set of concepts,
we form the analogue of the structural cognitive network, which only reflects the overlap of concepts; however, the
cognitive analogue of the functional networks in the probe can also be considered. Namely, there are linguistic cognitive networks, cognitive networks of associations, etc. Each concept can belong to several cognitive networks similar to the
multi-specialized neurons on the probe. Hence, we should assign the index $i$ to concepts that run in all cognitive networks to which the concept belongs.

For adults in consciousness, we can investigate the possible phase transitions in the new cognitive networks.
The standard phase transitions in the networks involve
clustering, localization of the excitations, percolation of the bonds, and clicks. 
A discussion of some critical phenomena in cognitive networks can be found in \cite{siew2019cognitive,stella2022cognitive}.
As an example of the percolation phenomenon, we can mention the critical size of the
percolating clusters in cognitive networks \cite{Gorsky7+-2}. This is the target
space analogue of the $7\pm2$ magic law for the working memory capacity. It
resembles the similar mechanism discussed in
\cite{Rabinovich2009} for the probe. It is not completely clear
what the meaning is of the possible localization phase transition in cognitive networks.

\section{Transitions in the probe networks. Order parameters}

Let us discuss phase transitions on the probe. The general idea is not new - we expect to observe some level of synchronization of neurons, or more precisely the synchronization of the functional networks. There are at least three different synchronization regimes.
\begin{itemize}
    \item Synchronization of the functional networks charged with respect to the mental part of the target 
    manifold at sleep. There is the continuous monitoring of the mental part of the target at sleep which
    implies the synchronization of the corresponding functional networks in the probe.
    
What could be an order parameter for  synchronization
in the real brain at sleep?  We have the time series from the FMRT data, and there is a suggestion \cite{tononi2003measuring}
to investigate the correlation of clusters in the effective networks built from the time series. Most efficiently,
this is done through the spectral analysis of the effective network, since the clusters correspond to the isolated eigenvalues \cite{toker2019information}. In any case, from fairly sophisticated
analysis of cluster correlations in time series, an order parameter $\Phi$ was invented that works reasonably well for patients in coma.

    \item Additional synchronization of the functional networks charged with respect to the material stimuli 
    at awaking transition. Mutual synchronization of two groups of the functional networks.

   In the awakening transition, we assume global synchronization of the network ensemble when the brain as a whole starts to operate in a new frequency band. The mechanisms of synchronization can be based on the physical interaction of some individual neurons of functional networks. This could be described by the oversimplified Kuramoto model of the multilayered networks where some sites belong
to the several layers corresponding to multi-specialized neurons. At some critical value of the
interaction and noise, synchronization could take place.  Another mechanism discussed in \cite{gorsky2022page}
is based on effective entropic forces. The synchronization transition on the probe can be considered as the integration
of components that occurs at the Page transition in the entangled system \cite{page1993average,page1993information}.

The order parameter for the awakening transition suggested in \cite{gorsky2022page} is based on the idea of the Page transition for entangled subsystems. It was assumed that the state of a brain and its EM rhythms are in an entangled state. This implies, for instance, that during the time from sleep to awakening the entropy evaluated from the time series has an upper limit and there exists the so called Page time
when the slope of the entropy of the time series gets changed:
\begin{equation}
    \frac{dS}{dt}>0 \quad \rightarrow \quad \frac{dS}{dt}\leq 0 
\end{equation}
Here, the sign of the entropy change can serve as the order parameter. This can be tested by an experiment during the awakening process and in anesthesia exit.

    \item Additional synchronization of the several functional networks during the measuring process(monitoring). This corresponds
    to the formation of the chimera states on the top of overall level synchronization.

For example, when we see the "big red apple", three functional networks, "big", "red" and "apple", get synchronized. It can
also be considered as the kind of phase transition well known from synchronization theory when the
chimera states get produced. In the chimera states only part of the degrees of freedom are synchronized.
The number of qualia emerged can serve as the order parameter for this second transition.
In rather formal terms, the order parameter here is the special parameter known in the framework of synchronization which indicates the presence of chimera states.

\end{itemize}

\begin{figure}
    \centering
    \includegraphics[width=0.55\linewidth]{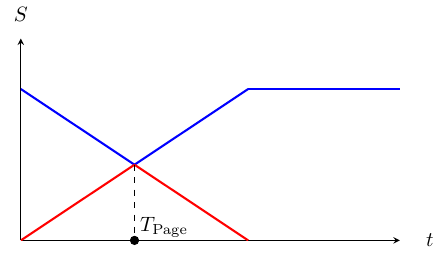}
    \caption{Page curve for the entropy of  time series of the brain activity(red)}
    \label{fig:picture}
\end{figure}

\begin{figure}
    \centering
    \includegraphics{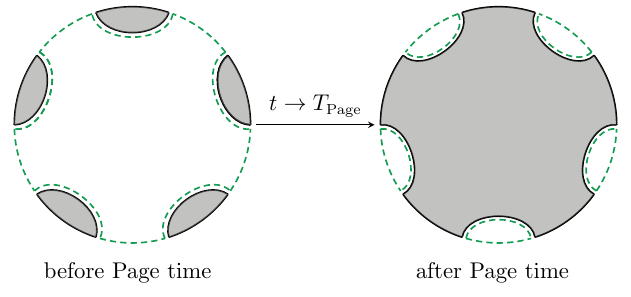}
    \caption{At the Page time
    the multiple "islands'-functional connectomes get consolidated into the disc-like geometry.}
    \label{fig:Islands_Page}
\end{figure}

Can we see the synchronization phase transition from the target space "observer"? To answer this question, let us recall some aspects of the so-called information geometry \cite{amari2016information}.
In a system dependent on some number of parameters, the probabilistic nature of evolution of the system induces the so-called information metric, or Fisher metric, on the parameter space, which in general exhibits the complicated geometry.
The phase transition in the system is believed to correspond to the singularity of the information metric in the parameter manifold \cite{kolodrubetz2013classifying}.
In our case, the target space plays the role of the parameter space from the viewpoint of the probe. 
The very precise example of such a singularity in the parameter space for the synchronization phase transition has been found in \cite{alexandrov2023information}
for the Kuramoto model with noise. 

We can conjecture that the special geodesic motion in the target space of information metrics takes place for partial synchronization towards some local minima in effective potential. 
In the "apple example" above, the metric components $g_{\mu\nu}$ in three directions $(\mu,\nu \in {\text
"big-red-apple"})$  in the target manifold are relevant.
We emphasize again that these synchronization transitions occur on the probe and that the target manifold of stimuli plays the role of the parameter space.


\section{Conclusion}

In this paper, we take advantage of a few old and relatively new ideas and focus on two main points. First,
we argue that the "probe-target" picture underlying the holographic-like approach is useful.
Representing both probe and target as ensembles of functional and cognitive networks, respectively, we can use the proper ideas from the theory of critical phenomena in out-of-equilibrium systems to describe the possible phase transitions
or crossovers both in the probe and in the target networks. The second point which seems to be new is the suggestion of identifying the "newborn-adult"   transition in the mental part of the target monitored by an
ensemble of functional networks as a version of the MIPT.

Two aspects of MIPT are of greatest importance. First, frequent monitoring at intermediate strength
induces an attractive interaction between the constituents and the corresponding clusterization. This
effect of monitoring provides the generation of a single concept from the ensemble of qualia of a
similar context. This is a short-term process.
Secondly, considering the very MIPT as the learnability transition, we fit with the natural
requirement that the recollection of the initial qualia be available. Above the critical measurement
rate in the disentangled phase , we can get the initial information of the ensemble of qualia.

No doubts the picture we suggested captures only a very small fragment of the problem.
Our aim was to get some fruits from the "probe-target" picture of two
ensembles of the networks with mutual influence and to take advantage of the possible transitions available in this extremely oversimplified setup focusing at the importance of the MIPT in the formation of consciousness. Nevertheless, let us mention a few research directions, which certainly are related with our study of MIPT.
Since we consider generative models in growing networks, network renormalization tools can be applied \cite{gabrielli2024network}. We have mentioned space-like entanglement as the order parameter for the MIPT transition; however, the time-like entanglement entropy of the different time intervals recently introduced in \cite{harper2023timelike,foligno2023temporal,milekhin2025observablecomputableentanglementtime} certainly can serve as another order parameter. Since we suggested recall as an act of measurement, the underlying mechanisms of associative memory such as
in the energy-based  Hopfield model \cite{hopfield1982neural,amari1972learning,krotov2023new} must be taken into account. In particular, recent non-trivial findings in the diffusive models of the memorization- generalization transition \cite{pham2025memorizationgeneralizationemergencediffusion} can be of some importance in our framework.

We certainly recognize the disputable nature of our paper. It can be criticized from the biological, physical, and philosophical viewpoints. Nevertheless, we believe that our attempt to apply the theory of transitions in out-of-equilibrium 
monitored systems to a
generative model for formation of consciousness is useful for this research area. We hope to develop a more formal
description of the generative model within the suggested framework. 

The author thanks K. Anokhin, A. Alexandrov and especially B. Rusakov for discussions and comments on the manuscript. The author is also grateful to A. Milekhin and F. Popov for some explanations concerning MIPT.

\section{Appendix}

Let us briefly comment on the few notions used in the text.
\begin{itemize}
\item{\bf Order parameter for general MIPT}
In the text  we have noted that the entropy of the measurement outcome serves as the order parameter
for the MIPT. Let us explain how it can be defined for the unitary circuit with fixed number 
of unitary gates and fixed positions and times of the measurement. Introduce the following matrix
\begin{equation}
  N_t(\rho)= \sum_m K_m\rho K^+_m  
  \label{sum}
\end{equation}
where $\rho$ is the system's density matrix and $K_m$ is the chain of the Kraus operators
\begin{equation}
    K_m(t)=K_t^{m_t} K_{t-1}^{m_t-1} \dots K_1^{m_1},\qquad K_t^{m_t}=P_t^{m_t}U_t
\end{equation}
The $U_t$ is unitary gate while $P_t^{m_t}$ is the random projector onto measurement outcome $m_t$. 

The (\ref{sum}) can be interpreted as a sum over trajectories in the space of states and 
\begin{equation}
    Tr K_{m_t}\rho K^+_{m_t}=p_{m_t}(\rho)
\end{equation}
is the probability of the the outcome $m_t$ which does not depend on the $\rho$ at the criticality.

It it useful to represent the (\ref{sum}) as the ensemble of the statistical models in the following
way \cite{li2021conformal}. Denote the partition function of the (1+1)-dimensional statistical model via the relation $$Z_m=p_m$$
hence we obtain the ensemble of models with quenched spacetime randomness. One can represent the partition function as 
\begin{equation}
    Z_m=\sum_i(\sigma_i^m)^2
\end{equation}
where $(\sigma_i^m)^2$ are the eigenvalues of the operator $K_mK_m^+$. At $t\rightarrow \infty$ the Lyapunov exponents
can be defined as 
\begin{equation}
   (\sigma_i^m)^2(t)=e^{\lambda_i^m t} 
\end{equation}
and for the leading Lyapunov exponent we obtain
\begin{equation}
    t \lambda_0^m =\ln p_m
\end{equation}
Hence remarkably the averaged free energy of the statistical ensemble $F$ is the Shannon entropy of the
measurement outcome
\begin{equation}
    F=-\sum_mp_m\ln p_m
\end{equation}
and serves as the order parameter for MIPT.

\item {\bf Toy measurement protocol}.
In our study we used the phenomenon of the measurement induced  attractive interaction between constituents \cite{doggen2022generalized} at the
particular domain in the 2-dimensional parameter space. Let us following
\cite{doggen2022generalized} present the toy measurement protocol involving
nonunitary measurement Hamiltonian 

\begin{equation}
    H_{meas}^{j}= iM\sum_{x} p_{x}^isgn(n_x-m_x^j)\hat{n}_{x}
\end{equation}
where $n_x$ is the expectation value of the onsite density, $p_x^j$ is binary variable, $m_x^j$ is random variable uniformly distributed 
in $(0,1)$, M is real and positive.  The measurement involves quench by adding $H_{meas}$ at time moment $t^i$ and 
removing at time $t^{i+1}$ and corresponding renormalization of probability. 

The variable $p_x^i$ indicates that measurement is performed at site $x$ at time $t^i$ with probability $P_x^i$ . If 
probability is uniform in space and time $P_x^i=P$ it parametrizes the measurement rate. The variable M parametrizes the
strength of the measurement and at $MT>>1$ we have projective measurement while the opposite limit corresponds to the weak 
measurement. Hence we have natural two dimensional parameter space $(M,P)$ mentioned in the text. It was argued in \cite{doggen2022generalized}
that such protocol is consistent with the more conventional description of measurement by the Kraus operators. 
In the indirect way the whole evolution can be thought of as the combination of the periods of the real-time 
evolution with the unitary Hamiltonian and
Euclidean evolution with the measurement Hamiltonian.

We conjecture in this study that the system under measurement is the ensemble of qualia. The measurement is considered as the
recollection of the i-th qualia that is backward-forward process. The analogue of the measurement outcome is weather the qualia is recollected or not
with some probability. After the recollection the whole ensemble of qualia is renormalized. The analogue of two dimensional
parameter space $(M,P)$ is natural. M measures the " intensity of recollection" while P measures its frequency.

\item{\bf On the toy "Hamiltonian"  of  a probe}.
Suppose that the ensemble of the functional networks in the probe  has been formed. 
The effective description of the probe should take into account two points. First, the interaction
between neurons should capture the information from the structural connectome. Secondly, the
neuron specialization should be taken into description. This means the the neuron specialized at 
discrete stimulus $\alpha$ in the target network should fire at this point of the target. In the case of place cells or
grid cells when the part of the target is continuous the neuron fires not at a point but in some domain.
In general one could expect that the metric in the space part of the target and the adjacency matrix 
of the semantic memory network are involved into the effective description of the ensemble of 
functional networks. It should be a generalization of the Ising type Hamiltonian 
used to describe the system of neurons without taking into account their specializations \cite{schneidman2006weak,toulouse1986spin}.

The specialization of neurons can be taken into account by the rectangular $N\times M$ "charge matrix" B
whose element $B_{\alpha,i}=1$ if i-th neuron is charged with respect to the $\alpha$-th stimulus or zero otherwise.
The simplest toy model which mimics the mapping of the 
probe into target stimuli space is represented by the generalization of Ising type Hamiltonian:
\begin{equation}
    H=\sum_{ij}  A_{ij} S_{i} S_{j} + \sum_{i\alpha} T_{\alpha}(i) B_{\alpha,i} S_{i}
\end{equation}
$S_{i}$ is the spin-like variable attributed to the $i$-th neuron $i=1,\dots N$, $\alpha= 1,\dots M$, where $N$ 
is the number of neurons in the structural connectome and $M$ is the number of sites in the target cognitive network.  
$A_{ij}$ - weighted adjacency matrix of the structural connectome which reflects their correlations. 
$T_{\alpha}(i)$ - effective external "magnetic" field which quantifies  the back-reaction of  the stimuli on the  -i-th neuron. 

Note that charge matrix is time dependent $B(t)$ since neurons can change their specialization. The target space variables
$T_{\alpha}$ play the role of parameters in the toy Hamiltonian. Hence considering the probabilistic variables in the large N
spin chain the Fisher information metric $\Psi_{\alpha\beta}$ can be induced in the target  which quantifies the correlation between 
the target sites.

\end{itemize}
\bibliographystyle{unsrt}
\bibliography{references}

\begin{thebibliography}{10}

\bibitem{nartallo2025nonequilibrium}
Ram{\'o}n Nartallo-Kaluarachchi, Morten~L Kringelbach, Gustavo Deco, Renaud Lambiotte, and Alain Goriely.
\newblock Nonequilibrium physics of brain dynamics.
\newblock {\em arXiv preprint arXiv:2504.12188}, 2025.

\bibitem{mora2011biological}
Thierry Mora and William Bialek.
\newblock Are biological systems poised at criticality?
\newblock {\em Journal of Statistical Physics}, 144:268--302, 2011.

\bibitem{bak}
P.~Bak.
\newblock {\em How Nature Works: The Science of Self-Organised Criticality}.
\newblock Copernicus Press, New York, 1996.

\bibitem{pikovsky2001synchronization}
Arkady Pikovsky, Michael Rosenblum, and J{\"u}rgen Kurths.
\newblock Synchronization.
\newblock {\em Cambridge university press}, 12, 2001.

\bibitem{buzsaki2006rhythms}
Gy{\"o}rgy Buzsaki.
\newblock Rhythms of the brain, 2006.

\bibitem{munoz2018colloquium}
Miguel~A Munoz.
\newblock Colloquium: Criticality and dynamical scaling in living systems.
\newblock {\em Reviews of Modern Physics}, 90(3):031001, 2018.

\bibitem{plenz2021self}
Dietmar Plenz, Tiago~L Ribeiro, Stephanie~R Miller, Patrick~A Kells, Ali Vakili, and Elliott~L Capek.
\newblock Self-organized criticality in the brain.
\newblock {\em Frontiers in Physics}, 9:639389, 2021.

\bibitem{tegmark2015consciousness}
Max Tegmark.
\newblock Consciousness as a state of matter.
\newblock {\em Chaos, Solitons \& Fractals}, 76:238--270, 2015.

\bibitem{kozma2018phase}
Robert Kozma and Joshua~JJ Davis.
\newblock Why do phase transitions matter in minds?
\newblock {\em Journal of Consciousness Studies}, 25(1-2):131--150, 2018.

\bibitem{chialvo2010emergent}
Dante~R Chialvo.
\newblock Emergent complex neural dynamics.
\newblock {\em Nature physics}, 6(10):744--750, 2010.

\bibitem{crick2003framework}
Francis Crick and Christof Koch.
\newblock A framework for consciousness.
\newblock {\em Nature neuroscience}, 6(2):119--126, 2003.

\bibitem{hameroff2014consciousness}
Stuart Hameroff and Roger Penrose.
\newblock Consciousness in the universe: A review of the ‘orch or’theory.
\newblock {\em Physics of life reviews}, 11(1):39--78, 2014.

\bibitem{tononi2004information}
Giulio Tononi.
\newblock An information integration theory of consciousness.
\newblock {\em BMC neuroscience}, 5(1):1--22, 2004.

\bibitem{tononi2015consciousness}
Giulio Tononi and Christof Koch.
\newblock Consciousness: here, there and everywhere?
\newblock {\em Philosophical Transactions of the Royal Society B: Biological Sciences}, 370(1668):20140167, 2015.

\bibitem{dehaene2001towards}
Stanislas Dehaene and Lionel Naccache.
\newblock Towards a cognitive neuroscience of consciousness: basic evidence and a workspace framework.
\newblock {\em Cognition}, 79(1-2):1--37, 2001.

\bibitem{seth2022theories}
Anil~K Seth and Tim Bayne.
\newblock Theories of consciousness.
\newblock {\em Nature reviews neuroscience}, 23(7):439--452, 2022.

\bibitem{bayne2023consciousness}
Tim Bayne, Joel Frohlich, Rhodri Cusack, Julia Moser, and Lorina Naci.
\newblock Consciousness in the cradle: on the emergence of infant experience.
\newblock {\em Trends in cognitive sciences}, 27(12):1135--1149, 2023.

\bibitem{li2018quantum}
Yaodong Li, Xiao Chen, and Matthew~PA Fisher.
\newblock Quantum zeno effect and the many-body entanglement transition.
\newblock {\em Physical Review B}, 98(20):205136, 2018.

\bibitem{skinner2019measurement}
Brian Skinner, Jonathan Ruhman, and Adam Nahum.
\newblock Measurement-induced phase transitions in the dynamics of entanglement.
\newblock {\em Physical Review X}, 9(3):031009, 2019.

\bibitem{chan2019unitary}
Amos Chan, Rahul~M Nandkishore, Michael Pretko, and Graeme Smith.
\newblock Unitary-projective entanglement dynamics.
\newblock {\em Physical Review B}, 99(22):224307, 2019.

\bibitem{fisher2023random}
Matthew~PA Fisher, Vedika Khemani, Adam Nahum, and Sagar Vijay.
\newblock Random quantum circuits.
\newblock {\em Annual Review of Condensed Matter Physics}, 14(1):335--379, 2023.

\bibitem{Barratt_2022}
Fergus Barratt, Utkarsh Agrawal, Andrew~C. Potter, Sarang Gopalakrishnan, and Romain Vasseur.
\newblock Transitions in the learnability of global charges from local measurements.
\newblock {\em Physical Review Letters}, 129(20), November 2022.

\bibitem{Ippoliti}
Matteo Ippoliti and Vedika Khemani.
\newblock Learnability transitions in monitored quantum dynamics via eavesdropper’s classical shadows.
\newblock {\em PRX}, 5(2):020304, 2024.

\bibitem{agrawal2024observing}
Utkarsh Agrawal, Javier Lopez-Piqueres, Romain Vasseur, Sarang Gopalakrishnan, and Andrew~C Potter.
\newblock Observing quantum measurement collapse as a learnability phase transition.
\newblock {\em Physical Review X}, 14(4):041012, 2024.

\bibitem{PhysRevLett.125.030505}
Soonwon Choi, Yimu Bao, Xiao-Liang Qi, and Ehud Altman.
\newblock Quantum error correction in scrambling dynamics and measurement-induced phase transition.
\newblock {\em Phys. Rev. Lett.}, 125:030505, Jul 2020.

\bibitem{szyniszewski2019entanglement}
Marcin Szyniszewski, Alessandro Romito, and Henning Schomerus.
\newblock Entanglement transition from variable-strength weak measurements.
\newblock {\em Physical Review B}, 100(6):064204, 2019.

\bibitem{szyniszewski2020universality}
Marcin Szyniszewski, Alessandro Romito, and Henning Schomerus.
\newblock Universality of entanglement transitions from stroboscopic to continuous measurements.
\newblock {\em Physical review letters}, 125(21):210602, 2020.

\bibitem{doggen2022generalized}
Elmer~VH Doggen, Yuval Gefen, Igor~V Gornyi, Alexander~D Mirlin, and Dmitry~G Polyakov.
\newblock Generalized quantum measurements with matrix product states: Entanglement phase transition and clusterization.
\newblock {\em Physical Review Research}, 4(2):023146, 2022.

\bibitem{hooft1993dimensional}
Gerard’t Hooft.
\newblock Dimensional reduction in quantum gravity.
\newblock {\em arXiv preprint gr-qc/9310026}, 1993.

\bibitem{susskind1995world}
Leonard Susskind.
\newblock The world as a hologram.
\newblock {\em Journal of Mathematical Physics}, 36(11):6377--6396, 1995.

\bibitem{lewis1956mind}
Clarence~Irving Lewis.
\newblock {\em Mind and the world-order: Outline of a theory of knowledge}.
\newblock Courier Corporation, 1956.

\bibitem{rusakov-concepts}
Boris Rusakov.
\newblock Concepts as elementary constituents of human consciousness.
\newblock {\em arXiv: 2208.09290}, Jul-Oct 2022.

\bibitem{Tegmark_2000}
Max Tegmark.
\newblock Importance of quantum decoherence in brain processes.
\newblock {\em Physical Review E}, 61(4):4194–4206, April 2000.

\bibitem{anokhin2021cognitome}
KV~Anokhin.
\newblock The cognitome: Seeking the fundamental neuroscience of a theory of consciousness.
\newblock {\em Neuroscience and Behavioral Physiology}, 51(7):915--937, 2021.

\bibitem{resende2022qualia}
Pedro Resende.
\newblock Qualia as physical measurements: a mathematical model of qualia and pure concepts.
\newblock {\em arXiv preprint arXiv:2203.10602}, 2022.

\bibitem{sporns2016networks}
Olaf Sporns.
\newblock {\em Networks of the Brain}.
\newblock MIT press, 2016.

\bibitem{lynn2019physics}
Christopher~W Lynn and Danielle~S Bassett.
\newblock The physics of brain network structure, function and control.
\newblock {\em Nature Reviews Physics}, 1(5):318--332, 2019.

\bibitem{moser2008place}
Edvard~I Moser, Emilio Kropff, and May-Britt Moser.
\newblock Place cells, grid cells, and the brain's spatial representation system.
\newblock {\em Annu. Rev. Neurosci.}, 31(1):69--89, 2008.

\bibitem{bowers2009biological}
Jeffrey~S Bowers.
\newblock On the biological plausibility of grandmother cells: implications for neural network theories in psychology and neuroscience.
\newblock {\em Psychological review}, 116(1):220, 2009.

\bibitem{rey2025lack}
Hernan~G Rey, Theofanis~I Panagiotaropoulos, Lorenzo Gutierrez, Fernando~J Chaure, Alejandro Nasimbera, Santiago Cordisco, Fabian Nishida, Antonio Valentin, Gonzalo Alarcon, Mark~P Richardson, et~al.
\newblock Lack of context modulation in human single neuron responses in the medial temporal lobe.
\newblock {\em Cell Reports}, 44(1), 2025.

\bibitem{rusakov-24}
Boris Rusakov.
\newblock Mathematical philosophy of consciousness.
\newblock {\em Journal of Consciousness Exploration and Research}, 15(2):179--189, 2024.

\bibitem{Saumier}
D.~Saumier and H~Chertkow.
\newblock Sematic memory.
\newblock {\em Curr Neurol Neurosci Rep}, 2(20), November 2002.

\bibitem{Tulving_1984}
Endel Tulving.
\newblock Précis of elements of episodic memory.
\newblock {\em Behavioral and Brain Sciences}, 7(2):223–238, 1984.

\bibitem{buzsaki2017space}
Gy{\"o}rgy Buzs{\'a}ki and Rodolfo Llin{\'a}s.
\newblock Space and time in the brain.
\newblock {\em Science}, 358(6362):482--485, 2017.

\bibitem{ginosar2023grid}
Gily Ginosar, Johnatan Aljadeff, Liora Las, Dori Derdikman, and Nachum Ulanovsky.
\newblock Are grid cells used for navigation? on local metrics, subjective spaces, and black holes.
\newblock {\em Neuron}, 111(12):1858--1875, 2023.

\bibitem{gorsky2022page}
Alexander Gorsky.
\newblock Page time and the order parameter for a consciousness state.
\newblock {\em arXiv preprint arXiv:2212.10602}, 2022.

\bibitem{pospelov2019spectral}
Nikita Pospelov, S~Nechaev, K~Anokhin, O~Valba, V~Avetisov, and A~Gorsky.
\newblock Spectral peculiarity and criticality of a human connectome.
\newblock {\em Physics of life reviews}, 31:240--256, 2019.

\bibitem{bobyleva2025metric}
Anna Bobyleva, Alexander Gorsky, Sergei Nechaev, Olga Valba, and Nikita Pospelov.
\newblock Metric structural human connectomes: localization and multifractality of eigenmodes.
\newblock {\em Network Neuroscience}, 9(2):682--711, 2025.

\bibitem{gullans2020dynamical}
Michael~J Gullans and David~A Huse.
\newblock Dynamical purification phase transition induced by quantum measurements.
\newblock {\em Physical Review X}, 10(4):041020, 2020.

\bibitem{choi2020quantum}
Soonwon Choi, Yimu Bao, Xiao-Liang Qi, and Ehud Altman.
\newblock Quantum error correction in scrambling dynamics and measurement-induced phase transition.
\newblock {\em Physical Review Letters}, 125(3):030505, 2020.

\bibitem{fan2021self}
Ruihua Fan, Sagar Vijay, Ashvin Vishwanath, and Yi-Zhuang You.
\newblock Self-organized error correction in random unitary circuits with measurement.
\newblock {\em Physical Review B}, 103(17):174309, 2021.

\bibitem{zhou2020entanglement}
Tianci Zhou and Adam Nahum.
\newblock Entanglement membrane in chaotic many-body systems.
\newblock {\em Physical Review X}, 10(3):031066, 2020.

\bibitem{jian2020measurement}
Chao-Ming Jian, Yi-Zhuang You, Romain Vasseur, and Andreas~WW Ludwig.
\newblock Measurement-induced criticality in random quantum circuits.
\newblock {\em Physical Review B}, 101(10):104302, 2020.

\bibitem{bao2020theory}
Yimu Bao, Soonwon Choi, and Ehud Altman.
\newblock Theory of the phase transition in random unitary circuits with measurements.
\newblock {\em Physical Review B}, 101(10):104301, 2020.

\bibitem{heyl2018dynamical}
Markus Heyl.
\newblock Dynamical quantum phase transitions: a review.
\newblock {\em Reports on Progress in Physics}, 81(5):054001, 2018.

\bibitem{zvyagin2016dynamical}
AA~Zvyagin.
\newblock Dynamical quantum phase transitions.
\newblock {\em Low Temperature Physics}, 42(11):971--994, 2016.

\bibitem{pütz2025flownishimoriuniversalityweakly}
Malte Pütz, Romain Vasseur, Andreas W.~W. Ludwig, Simon Trebst, and Guo-Yi Zhu.
\newblock Flow to nishimori universality in weakly monitored quantum circuits with qubit loss, 2025.

\bibitem{zabalo2022operator}
Aidan Zabalo, Michael~J Gullans, Justin~H Wilson, Romain Vasseur, Andreas~WW Ludwig, Sarang Gopalakrishnan, David~A Huse, and JH~Pixley.
\newblock Operator scaling dimensions and multifractality at measurement-induced transitions.
\newblock {\em Physical review letters}, 128(5):050602, 2022.

\bibitem{friston2010free}
Karl Friston.
\newblock The free-energy principle: a unified brain theory?
\newblock {\em Nature reviews neuroscience}, 11(2):127--138, 2010.

\bibitem{dacosta2024mathematicalperspectiveneurophenomenology}
Lancelot~Da Costa, Lars Sandved-Smith, Karl Friston, Maxwell J.~D. Ramstead, and Anil~K. Seth.
\newblock A mathematical perspective on neurophenomenology, 2024.

\bibitem{whyte2024minimaltheoryconsciousnessimplicit}
Christopher~J. Whyte, Andrew~W. Corcoran, Jonathan Robinson, Ryan Smith, Rosalyn~J. Moran, Thomas Parr, Karl~J. Friston, Anil~K. Seth, and Jakob Hohwy.
\newblock On the minimal theory of consciousness implicit in active inference, 2024.

\bibitem{siew2019cognitive}
Cynthia~SQ Siew, Dirk~U Wulff, Nicole~M Beckage, and Yoed~N Kenett.
\newblock Cognitive network science: A review of research on cognition through the lens of network representations, processes, and dynamics.
\newblock {\em Complexity}, 2019(1):2108423, 2019.

\bibitem{stella2022cognitive}
Massimo Stella.
\newblock Cognitive network science for understanding online social cognitions: A brief review.
\newblock {\em Topics in Cognitive Science}, 14(1):143--162, 2022.

\bibitem{Gorsky7+-2}
Olga Valba and Alexander Gorsky.
\newblock K‑clique percolation in free association networks and the possible mechanism behind the 7 ± 2 law.
\newblock {\em Scientific Reports}, 12:5540, 2022.

\bibitem{Rabinovich2009}
Christian Bick and Mikhail Rabinovich.
\newblock Dynamical origin of the effective storage capacity in the brain’s working memory.
\newblock {\em Phys. Rev. Letters}, 103(21):218101(1--4), 2009.

\bibitem{tononi2003measuring}
Giulio Tononi and Olaf Sporns.
\newblock Measuring information integration.
\newblock {\em BMC neuroscience}, 4(1):1--20, 2003.

\bibitem{toker2019information}
Daniel Toker and Friedrich~T Sommer.
\newblock Information integration in large brain networks.
\newblock {\em PLoS computational biology}, 15(2):e1006807, 2019.

\bibitem{page1993average}
Don~N Page.
\newblock Average entropy of a subsystem.
\newblock {\em Physical review letters}, 71(9):1291, 1993.

\bibitem{page1993information}
Don~N Page.
\newblock Information in black hole radiation.
\newblock {\em Physical review letters}, 71(23):3743, 1993.

\bibitem{amari2016information}
Shun-ichi Amari.
\newblock {\em Information geometry and its applications}, volume 194.
\newblock Springer, 2016.

\bibitem{kolodrubetz2013classifying}
Michael Kolodrubetz, Vladimir Gritsev, and Anatoli Polkovnikov.
\newblock Classifying and measuring geometry of a quantum ground state manifold.
\newblock {\em Physical Review B—Condensed Matter and Materials Physics}, 88(6):064304, 2013.

\bibitem{alexandrov2023information}
Artem Alexandrov and Alexander Gorsky.
\newblock Information geometry and synchronization phase transition in the kuramoto model.
\newblock {\em Physical Review E}, 107(4):044211, 2023.

\bibitem{gabrielli2024network}
Andrea Gabrielli, Diego Garlaschelli, Subodh~P Patil, and M~Serrano.
\newblock Network renormalization.
\newblock {\em arXiv preprint arXiv:2412.12988}, 2024.

\bibitem{harper2023timelike}
Jonathan Harper, Ali Mollabashi, Tadashi Takayanagi, Yusuke Taki, et~al.
\newblock Timelike entanglement entropy.
\newblock {\em Journal of High Energy Physics}, 2023(5):1--62, 2023.

\bibitem{foligno2023temporal}
Alessandro Foligno, Tianci Zhou, and Bruno Bertini.
\newblock Temporal entanglement in chaotic quantum circuits.
\newblock {\em Physical Review X}, 13(4):041008, 2023.

\bibitem{milekhin2025observablecomputableentanglementtime}
Alexey Milekhin, Zofia Adamska, and John Preskill.
\newblock Observable and computable entanglement in time, 2025.

\bibitem{hopfield1982neural}
John~J Hopfield.
\newblock Neural networks and physical systems with emergent collective computational abilities.
\newblock {\em Proceedings of the national academy of sciences}, 79(8):2554--2558, 1982.

\bibitem{amari1972learning}
S-I Amari.
\newblock Learning patterns and pattern sequences by self-organizing nets of threshold elements.
\newblock {\em IEEE Transactions on computers}, 100(11):1197--1206, 1972.

\bibitem{krotov2023new}
Dmitry Krotov.
\newblock A new frontier for hopfield networks.
\newblock {\em Nature Reviews Physics}, 5(7):366--367, 2023.

\bibitem{pham2025memorizationgeneralizationemergencediffusion}
Bao Pham, Gabriel Raya, Matteo Negri, Mohammed~J. Zaki, Luca Ambrogioni, and Dmitry Krotov.
\newblock Memorization to generalization: Emergence of diffusion models from associative memory, 2025.

\bibitem{li2021conformal}
Yaodong Li, Xiao Chen, Andreas~WW Ludwig, and Matthew~PA Fisher.
\newblock Conformal invariance and quantum nonlocality in critical hybrid circuits.
\newblock {\em Physical Review B}, 104(10):104305, 2021.

\bibitem{schneidman2006weak}
Elad Schneidman, Michael~J Berry, Ronen Segev, and William Bialek.
\newblock Weak pairwise correlations imply strongly correlated network states in a neural population.
\newblock {\em Nature}, 440(7087):1007--1012, 2006.

\bibitem{toulouse1986spin}
G{\'e}rard Toulouse, Stanislas Dehaene, and Jean-Pierre Changeux.
\newblock Spin glass model of learning by selection.
\newblock {\em Proceedings of the National Academy of Sciences}, 83(6):1695--1698, 1986.

\end{thebibliography}

\end{document}